\documentclass[showpacs,nofootinbib]{revtex4}
\usepackage{amsfonts}
\usepackage{amsmath}
\usepackage{graphicx}
\usepackage{amssymb}

\input{tcilatex}
\begin{document}

\title{An Investigation of Equivalence between Bulk-based and Brane-based
Approaches}
\author{G\"{u}l\c{c}in Uluyazi}
\affiliation{Physics Department, \.{I}stanbul University, Vezneciler, Turkey.}

\begin{abstract}
There are two different approaches to handle brane-world models brane-based
or bulk-based. In the brane-based approach, the brane is chosen to be fixed
on a coordinate system, whereas in the bulk-based approach, it is no longer
static as it moves along the extra dimension. It is aimed to get a general
formalism of the equivalence between two approaches obtained for a specific
model in Mukohyama et all [12]. We found that calculations driven by a
general anisotropic bulk-based metric yields a brane-based metric in
Gaussian Normal Coordinates by conserving spatial anisotropy.
\end{abstract}

\date{12 april 2011}
\maketitle
\maketitle

\address{Physics Department, Istanbul University, \\
Istanbul, Turkey \\
$^*$E-mail: gberkoz@istanbul.edu.tr}

\section{Introduction}

The discovery of Rundall and Sundrum (RS) $[1,2]$ opened a new window to our
conventional cosmology. Their models give an alternative solution of
hierarchy problem in which the additional dimension does not have to be
small as in Kaluza-Klein model. The extra dimension can be infinite but
warped. Our observable universe could be a 1+3 surface (called a brane)
embedded 5-dimensional spacetime (called a bulk). Standard Model matters and
fields are confined on the brane while gravity is permitted to propagate the
bulk.

\ RS Models and subsequent generalization from a Minkowski brane to a
Friedmann-Robertson-Walker (FRW) brane $[3,4,5,6,7,8,9]$ were derived as
solutions in particular coordinates of the 5-dimensional Einstein equations,
together with the junction conditions at the Z2-symmetric brane.

\begin{equation}
R_{AB}^{5}-\frac{1}{2}g_{AB}^{5}=-\Lambda _{5}g_{AB}^{5}+\kappa
_{5}^{2}T_{AB}^{5}
\end{equation}%
\newline
where ${T_{AB}^{5}}$ represents any 5-dimensional energy-momentum of the
gravitational sector and it provides a conservation equation 
\begin{equation}
\nabla _{A}T^{AB}=0
\end{equation}%
The 5-dimensional curvature tensor has Weyl and Ricci parts 
\begin{eqnarray}
R_{ABCD}^{5} &=&C_{ABCD}^{5}+\frac{1}{3}%
[R_{AC}^{5}g_{BD}^{5}-R_{AD}^{5}g_{BC}^{5}+R_{BD}^{5}g_{AC}^{5}-R_{BC}^{5}g_{AD}^{5}]\ 
\notag \\
&&-\frac{1}{12}R^{5}[g_{AC}^{5}g_{BD}^{5}-g_{AD}^{5}g_{BC}^{5}]
\end{eqnarray}%
\newline
where ${C_{ABCD}^{5}}$ is Weyl Tensor.

\ Another way of deriving the brane world cosmological equations is to use
induced equations on the brane. Shiromizu-Maeda-Sasaki $[10]$ obtained
4-dimensional induced field equations by projecting 5-dimensional equations
on the brane and using Israel Junction Conditions. 
\begin{equation}
G_{AB}^{4}=-\Lambda _{4}h_{AB}+8\pi G_{5}\tau _{AB}+\kappa _{5}^{4}\pi
_{AB}^{5}-E_{AB}^{5}
\end{equation}%
\begin{equation}
\Lambda _{4}=\frac{1}{2}\kappa _{5}^{2}[\Lambda _{5}+\frac{1}{2}\kappa
_{5}^{2}\lambda ^{2}]
\end{equation}%
\begin{equation}
G_{5}=\frac{\kappa _{5}^{4}\lambda }{48\pi }
\end{equation}%
\begin{equation}
\pi _{AB}^{5}=-\frac{1}{4}\tau _{AC}{\tau _{B}}^{C}+\frac{1}{12}\tau \tau
_{AB}+\frac{1}{8}h_{AB}\tau _{CD}\tau ^{CD}-\frac{1}{24}h_{AB}\tau ^{2}
\end{equation}

These equations are also important to give us the connection with four and
five dimensional quantities. They also contain bulk effects on the brane. In
our previous study, $[11]$ when investigating an anisotropic model, four
dimensional induced field equations were employed. But in this study, the
calculations are derived by using five dimensional equations (1).

Except from these two ways of cosmological model construction, there are
also two more approaches: brane-based or bulk based. In the brane-based
approach, the brane is chosen to be fixed on a coordinate system. The 5d
metric depends on both time and extra coordinate. On the contrary, in the
bulk-based approach the brane is no longer static as it moves along the
extra dimension and the bulk metric has static form. It is shown that both
approaches are equivalent when homogeneity and isotropy are considered $[12]$%
. There is no existing study showing what kind of equality will be obtained
in the manner of different models, especially anisotropic models which we
are interested in. The main purpose of the study is to generalizing the
formulation for the afterwards getting the corresponding versions in
brane-based approach.

We first construct formulas providing the transformation of the bulk-based
approach to the brane-based one in Sec.2. It is shown that our results cover
the solutions in $[12]$ in which the equivalence metric of Scw-AdS spacetime
has been explored. At the end of the driven calculations, we find that the
equivalence of the most general anisotropic bulk-based metric is an
anisotropic metric in the form of Gaussian Normal Coordinates as expected.

Because the Gaussian Normal Coordinates are the most operational coordinates
which allow to induce the metric from 5-dimensional spacetime to
4-dimensional hypersurface, the final conclusion of the study is important
for obtaining correspondence of any models in Gaussian Normal Coordinates.
Apart from constituting general formalism, we get the solutions for a
bulk-based model named Gergely-Maartens metric, and construct corresponding
brane-based solution as an application in Sec.3.

\section{General Transformation from Bulk-based Metric to Brane-based one}

\bigskip The most general 5-dimensional anisotropic bulk-based metric
admitting non-zero Killing vector space is 
\begin{equation}
ds^{2}=-A_{0}(\hat{r})d\hat{t}^{2}+A_{ij}(\hat{r})d\hat{x}^{i}d\hat{x}%
^{j}+A_{4}(\hat{r})d\hat{r}^{2}\   \label{bulkmetric}
\end{equation}%
where $\hat{r}$ denotes the extra spatial coordinate and $i,j=1,2,3$ are
indices of 3-dimensional spacetime. The 4-dimensional brane corresponding
our universe moves along the extra dimension and is described by $(\tau
,x^{i})$ coordinates. Base vectors and one forms in the 5-dimensional and
4-dimensional spacetime respectively are in below. 
\begin{equation}
e_{A}\equiv \partial _{A}=\frac{\partial }{\partial \hat{x}_{A}}=(\partial _{%
\hat{t}},\partial _{\hat{x}^{i}},\partial _{\hat{r}})\text{ \ \ \ \ \ }%
\omega ^{A}\equiv d\hat{x}^{A}=(d\hat{t},d\hat{x}^{i},d\hat{r})
\end{equation}%
\begin{equation}
e_{\mu }\equiv \partial _{\mu }=\frac{\partial }{\partial \hat{x}_{\mu }}%
=(\partial _{\tau },\partial _{x^{i}})\text{ \ \ \ \ \ }\omega ^{\mu }\equiv
dx^{\mu }=(d\tau ,dx^{i})
\end{equation}%
Here we define $A=0..4$, $\mu =0..3.$ The brane represented by $\hat{r}=R(%
\hat{t})$ hypersurfaces can be induced on 4-dimensional spacetime via
following transformations. 
\begin{equation}
\begin{array}{c}
\hat{t}=T(\tau )\rightarrow d\hat{t}=\dot{T}d\tau \\ 
\hat{x}^{i}=x^{i}\rightarrow d\hat{x}^{i}=dx^{i} \\ 
\hat{r}=R(\tau )\rightarrow d\hat{r}=\dot{R}d\tau%
\end{array}%
\end{equation}

The induced metric on brane is then

\bigskip 
\begin{equation}
ds_{brane}^{2}=-(A_{0}\dot{T}^{2}-A_{4}\dot{R}^{2})d\tau ^{2}+A_{ij}(R(\tau
))dx^{i}dx^{j}
\end{equation}%
where we introduced cosmological time $\tau $ and cosmological scale factor $%
R(\tau )$. The dot denotes to derivative respect to $\tau $. Now we can
construct a vector space generating by tangent vectors of geodesics which
intersect with hypersurface $\hat{r}=R(\hat{t})$ perpendicularly 
\begin{equation}
u^{A}=e_{\tau }^{A}\partial _{A}=\dot{T}\partial _{\hat{t}}+\dot{R}\partial
_{\hat{r}}  \label{tangent1}
\end{equation}

We choose geodesics as spacelike and have zero $\hat{x}^{i}$-components to
provide a timelike hypersurface. The Killing field of bulk spacetime helps
us to find constants of motion along geodesics. 
\begin{equation}
g_{AB}u^{A}\xi ^{B}=-E
\end{equation}%
\begin{equation}
g_{AB}u^{A}u^{B}=1
\end{equation}%
where $E$ is an integration constant. Using tangent vector's components in (%
\ref{tangent1}), we obtain

\bigskip 
\begin{equation}
u^{A}=\left( \frac{E}{A_{0}},0,0,0,\mp \frac{A_{0}+E^{2}}{A_{0}A_{4}}\right)
\end{equation}

The trajectory of the geodesic is given by 
\begin{equation}
\frac{dx^{A}}{dw}=u^{A}
\end{equation}%
here $w$ is the affine parameter. Every points $P$ on the hypersurface
described by $(\tau ,x^{i})$ coordinates intersect perpendicularly with an
affinely parameterized geodesic. Hence we can describe $P$ via a new
coordinate set $(\tau ,x^{i},w)$, where the new coordinate $w$ is now an
extra spatial coordinate of $P$ and this system is called by brane-based
coordinates. One can easily construct brane-based metric from bulk-based one
by applying transformations: $\hat{r}=\hat{r}(\tau ,w)$, $\hat{t}=\hat{t}%
(\tau ,w)$ 
\begin{equation}
d\hat{t}=\left( \frac{\partial \hat{t}}{\partial \tau }\right) d\tau +\left( 
\frac{\partial \hat{t}}{\partial w}\right) dw=e_{\tau }^{\hat{t}}d\tau
+e_{w}^{\hat{t}}dw
\end{equation}%
\begin{equation}
d\hat{r}=\left( \frac{\partial \hat{r}}{\partial \tau }\right) d\tau +\left( 
\frac{\partial \hat{r}}{\partial w}\right) dw=e_{\tau }^{\hat{r}}d\tau
+e_{w}^{\hat{r}}dw
\end{equation}%
substituting them in (\ref{bulkmetric}) gives

\bigskip 
\begin{eqnarray}
ds^{2} &=&-\left[ A_{0}\left( \frac{\partial \hat{t}}{\partial \tau }\right)
^{2}-A_{4}\left( \frac{\partial \hat{r}}{\partial \tau }\right) ^{2}\right]
d\tau ^{2}+A_{ij}dx^{i}dx^{j}\   \notag \\
&&+\left[ -A_{0}\left( \frac{\partial \hat{t}}{\partial w}\right)
^{2}+A_{4}\left( \frac{\partial \hat{r}}{\partial w}\right) ^{2}\right]
dw^{2}\   \notag \\
&&+2\left[ -A_{0}\left( \frac{\partial \hat{t}}{\partial \tau }\right)
\left( \frac{\partial \hat{t}}{\partial w}\right) +A_{4}\left( \frac{%
\partial \hat{r}}{\partial \tau }\right) \left( \frac{\partial \hat{r}}{%
\partial w}\right) \right] d\tau dw
\end{eqnarray}

The final metric in (20) is a general form of brane-based metric which is
transformed from bulk-based metric and here $w$ is the affine parameter. We
need to find the exact forms of transformation coefficients denoted by
partial derivatives in (20). Replacing (16) in (17) we get two equations 
\begin{equation}
u^{\hat{t}}=e_{w}^{\hat{t}}=\frac{\partial \hat{t}}{\partial w}=\frac{E}{%
A_{0}}
\end{equation}
\begin{equation}
u^{\hat{r}}=e_{w}^{\hat{r}}=\frac{\partial \hat{r}}{\partial w}=\mp\sqrt{ 
\frac{A_{0}+E^{2}}{A_{0}A_{4}}}
\end{equation}
in which the latter one gives a integral relation between extra coordinates
of two approaches. In the case of components of bulk-based metric are known,
(26) can be solved exactly. 
\begin{equation}
\mp w+w_{0}(\tau )=\int \frac{d\hat{r}}{\sqrt{\frac{A_{0}+E^{2}}{A_{0}A_{4}}}%
}
\end{equation}

On the other hand, we get transverse coefficients in (21-22) from $%
dw/dx^{B}=g_{AB}u^{A}$ 
\begin{equation}
e_{\hat{t}w}=\frac{\partial w}{\partial \hat{t}}=-E
\end{equation}
\begin{equation}
e_{\hat{x}^{i}w}=\frac{\partial w}{\partial \hat{x}^{i}}=0
\end{equation}
\begin{equation}
e_{\hat{r}w}=\frac{\partial w}{\partial \hat{r}}=\pm \sqrt{\frac{A_{4}\left(
A_{0}+E^{2}\right) }{A_{0}}}
\end{equation}

The integrability condition $ddw=0$ is equivalent to 
\begin{equation}
\left( \frac{\partial w}{\partial \hat{t}}\right) d\hat{t}=-\left( \frac{%
\partial w}{\partial \hat{r}}\right) d\hat{r}
\end{equation}
and gives a ratio of coefficients 
\begin{equation}
\left( \frac{\partial \hat{t}}{\partial \tau }\right) /\left( \frac{\partial 
\hat{r}}{\partial \tau }\right) =-\left( \frac{\partial w}{\partial \hat{r}}%
\right) /\left( \frac{\partial w}{\partial \hat{t}}\right) =\pm \sqrt{\frac{%
A_{4}\left( A_{0}+E^{2}\right) }{A_{0}E^{2}}}
\end{equation}
The transverse ratio is then 
\begin{equation}
\left( \frac{\partial \tau }{\partial \hat{t}}\right) /\left( \frac{\partial
\tau }{\partial \hat{r}}\right) =-\left( \frac{\partial \hat{r}}{\partial w}%
\right) /\left( \frac{\partial \hat{t}}{\partial w}\right) =\pm \sqrt{\frac{%
A_{0}\left( A_{0}+E^{2}\right) }{A_{4}E^{2}}}
\end{equation}

In order to see all transformation coefficients which we have calculated by
now let us put them into a matrix form as below 
\begin{equation}
(\hat{t},\hat{x}^{i},\hat{r})\rightarrow (\tau ,x^{i},w)\text{ : }\left( 
\begin{array}{ccc}
e_{\tau }^{\hat{t}} & e_{x^{i}}^{\hat{t}} & e_{w}^{\hat{t}} \\ 
e_{\tau }^{\hat{x}^{i}} & e_{x^{i}}^{\hat{x}^{i}} & e_{w}^{\hat{x}^{i}} \\ 
e_{\tau }^{\hat{r}} & e_{x^{i}}^{\hat{r}} & e_{w}^{\hat{r}}%
\end{array}%
\right) =\left( 
\begin{array}{ccc}
a_{11} & 0 & \frac{E}{A_{0}} \\ 
0 & 1 & 0 \\ 
a_{31} & 0 & \pm \sqrt{\frac{\left( A_{0}+E^{2}\right) }{A_{4}A_{0}}}%
\end{array}%
\right)
\end{equation}%
\begin{equation}
(\tau ,x^{i},w)\rightarrow (\hat{t},\hat{x}^{i},\hat{r})\text{ : }\left( 
\begin{array}{ccc}
e_{\hat{t}\tau }^{{}} & e_{\hat{t}x^{i}}^{{}} & e_{\hat{t}w}^{{}} \\ 
e_{\hat{x}^{i}\tau }^{{}} & e_{\hat{x}^{i}x^{i}}^{{}} & e_{\hat{x}^{i}w}^{{}}
\\ 
e_{\hat{r}\tau }^{{}} & e_{\hat{r}x^{i}}^{{}} & e_{\hat{r}w}^{{}}%
\end{array}%
\right) =\left( 
\begin{array}{ccc}
a^{11} & 0 & -E \\ 
0 & 1 & 0 \\ 
a^{31} & 0 & \pm \sqrt{\frac{A_{4}\left( A_{0}+E^{2}\right) }{A_{0}}}%
\end{array}%
\right)
\end{equation}%
Note that two of the matrix components remain unknown, but they depend on
each other according to these equations 
\begin{equation}
e_{\tau }^{\hat{t}}/e_{\tau }^{\hat{r}}=a_{11}/a_{31}=\pm \sqrt{\frac{%
A_{4}\left( A_{0}+E^{2}\right) }{A_{0}E^{2}}}
\end{equation}

\bigskip 
\begin{equation}
e_{\hat{t}\tau }^{{}}/e_{\hat{r}\tau }^{{}}=a^{11}/a^{31}=\pm \sqrt{\frac{%
A_{0}\left( A_{0}+E^{2}\right) }{A_{4}E^{2}}}
\end{equation}%
Substituting partial derivatives from above matrix in (20), we find final
form of brane-based metric

\begin{equation}
ds^{2}=-\frac{A_{0}(\hat{r}(\tau ,w))A_{4}(\hat{r}(\tau ,w))}{E^{2}}\left( 
\frac{\partial \hat{r}}{\partial \tau }\right) ^{2}d\tau ^{2}\ +A_{ij}(\hat{r%
}(\tau ,w))dx^{i}dx^{j}+dw^{2}
\end{equation}%
which is a general form of brane-based metric in Gaussian Normal Coordinates
but contains spatial anisotropy as expected. If we choose metric components
as $A_{0}=f(\hat{r})$, $A_{4}=1/f(\hat{r})$ and isotropic 3-dimensional
metric, Eq. (34) readily yields the result in Ref. $[12]$.

All terms in (34) depend on brane coordinates $(\tau,w)$. The motion
constant $E$ of bulk geodesics would not be a constant anymore on the brane
and turns out to be $E=E(\tau)$. Besides that the term showing partial
derivative can be calculated from integral relation (23) after obtaining
explicit forms of metric components. For this purpose one must solve vacuum
Einstein Field Equations in the bulk and replace them with transformed
metric.

The general form of brane-based metric for anisotropic model given by (34)
produce the induced metric on brane by setting $w=0$. The result metric
represents 4-dimensional spacetime which determined by Egs.(4-7).

\section{\protect\bigskip Transformation for Gergely-Maartens Metric}

In this section we present an application regarding the transformation
introduced above. Gergely-Maartens (GM) metric [13] can be taken as static
5d bulk-based metric for starters. This metric corresponds a Non-SchAds bulk
spacetime with static Friedman brane. 
\begin{equation}
\Gamma ^{2}ds^{2}=-F^{2}(\hat{r},\epsilon )d\hat{t}^{2}+d\hat{\chi}^{2}+%
\mathcal{H}\left( \hat{\chi};\epsilon \right) \left( d\hat{\theta}^{2}+\sin
^{2}\hat{\theta}d\hat{\phi}^{2}\right) +d\hat{r}^{2}
\end{equation}%
\begin{equation}
\mathcal{H}\left( \hat{\chi};k\right) =\left\{ 
\begin{array}{c}
\sin \hat{\chi}~,\qquad \epsilon =1 \\ 
\quad \hat{\chi}~~~,\qquad \epsilon =0 \\ 
~\sinh \hat{\chi}~,\qquad \epsilon =-1%
\end{array}%
\right. ~.
\end{equation}%
where $\Gamma ;\kappa ^{2}\Lambda _{5}=3\epsilon \Gamma ^{2}$ gives the
magnitude of the cosmological constant and $\epsilon $ its sign. If $%
\epsilon =0$, then $\Gamma $ is a removable constant. 
\begin{equation}
G_{AB}=-\kappa ^{2}\Lambda _{5}g_{AB}
\end{equation}%
By solving (35) from the 5-dimensional vacuum Einstein equation, we find 
\begin{equation}
F(\hat{r},\epsilon )=\left\{ 
\begin{array}{c}
A\cos (\sqrt{2}\hat{r})+B\sin (\sqrt{2}\hat{r}),\qquad \epsilon =1 \\ 
\quad (A+B\sqrt{2}\hat{r})~,\qquad \epsilon =0 \\ 
~A\cosh (\sqrt{2}\hat{r})+B\sinh (\sqrt{2}\hat{r})~,\qquad \epsilon =-1%
\end{array}%
\right. ~.
\end{equation}%
here $A$ and $B$ are constants.

To get corresponding brane-based metric of GM, we first have to solve the
integration in (23) by substituting $A_{0}=F^{2}(\hat{r},\epsilon ),A_{4}=1$ 
\begin{equation}
\mp w+w_{0}(\tau )=\int \frac{d\hat{r}}{\sqrt{\frac{F^{2}+E^{2}}{F^{2}}}}
\end{equation}
Solutions are listed below for each case of $\epsilon$ 
\begin{equation}
\mp w+w_{0}(\tau )=\left\{ 
\begin{array}{c}
\frac{1}{2\sqrt{2}}\arcsin \left[ \frac{\sqrt{a^{2}x^{2}+cx+b^{2}}}{%
(c^{2}-4ab)|x|}2i(ax-b)(2ab-c)\right] ,\qquad \epsilon =1 \\ 
\quad \frac{\sqrt{(A+B\sqrt{2}\hat{r})^{2}~+E^{2}}}{\sqrt{2}B},\qquad
\epsilon =0 \\ 
~\frac{1}{2\sqrt{2}}\arcsin \left[ \frac{\sqrt{\alpha ^{2}y^{2}+\beta
y+\gamma ^{2}}}{(4\alpha \gamma -\beta ^{2})|y|}2i(\alpha y-\gamma )(2\alpha
\gamma -\beta )\right] ~,\qquad \epsilon =-1%
\end{array}%
\right.
\end{equation}
Here we defined $x=\exp (i2\sqrt{2}\hat{r}),a=\frac{A}{2}+\frac{B}{2i},b=%
\frac{A}{2}-\frac{B}{2i},c=2ab+E^{2}$ and $y=\exp (2\sqrt{2}\hat{r}),\alpha =%
\frac{A}{2}+\frac{B}{2},\gamma =\frac{A}{2}-\frac{B}{2},\beta =2\alpha
\gamma +E^{2}$

We show explicit calculation for the $\epsilon =0$ case only, but the other
cases can also be handled by the same calculations. 
\begin{equation}
\left( A+B\sqrt{2}\hat{r}\right) ^{2}+E^{2}=2B^{2}\left( \mp w+w_{0}\right)
^{2}
\end{equation}
To determine the constants $E$ and $w_0$, we use the fact that the geodesic
intersects with the hypersurface $\hat{r} = R(\hat{t} )$ perpendicularly at $%
\hat{t} =\hat{t_0}$ and that the affine parameter $w$ is zero on the
hypersurface. 
\begin{equation}
\left( A+B\sqrt{2}R\right) ^{2}+E^{2}=2B^{2}w_{0}^{2}
\end{equation}
Substituting $w_0$ into (41) yields 
\begin{equation}
\hat{r}=\frac{1}{\sqrt{2}B}\left\{ \left( \mp \sqrt{2}Bw+\sqrt{A+B\sqrt{2}%
R+E^{2}}\right) ^{2}-E^{2}\right\} ^{1/2}-\frac{A}{\sqrt{2}B}
\end{equation}

Now we can derive brane-based metric of (35) as following 
\begin{equation}
ds^{2}=-\Psi (\tau ,w)d\tau ^{2}+d\chi ^{2}+\mathcal{H}\left( \chi ;\epsilon
\right) \left( d\theta ^{2}+\sin ^{2}\theta d\phi ^{2}\right) +dw^{2}
\end{equation}%
where

\begin{equation}
\Psi (\tau ,w)=\left\{ \frac{(A+B\sqrt{2}R)}{HR}\frac{\left( \mp
w+w_{0}\right) }{w_{0}}+\left( \frac{\dot{H}}{H}+H\right) \frac{\mp w}{\sqrt{%
2}Bw_{0}}\right\} \sqrt{2B^{2}\left( \mp w+w_{0}\right) ^{2}-H^{2}R^{2}}
\end{equation}%
and $H=\dot{R}/R$ is the Hubble constant. If one sets the $w=0$ in (45), it
could be shown that 4-dimensional metric is equivalent to Kantoski-Sach
spacetime [14].

\section{Acknowledgements}

This study is some part of author's Ph.D. Thesis which is supported by
Istanbul University Research Projects with number 4290.

\end{document}